\documentclass[12pt, a4paper]{article}

\def\bea{\begin{eqnarray*}}
\def\eea{\end{eqnarray*}}
\def\bean{\begin{eqnarray}}
\def\eean{\end{eqnarray}}
\def\bi{\begin{itemize}}
\def\ei{\end{itemize}}

\newcommand{\dint}{d_\text{int}}
\newcommand{\dfin}{d_\text{fin}}
\newcommand{\km}{KM }
\newcommand{\kmns}{KM}

\let\code=\texttt

\usepackage{amsmath}
\usepackage{amssymb}
\usepackage{amsfonts}
\usepackage{amsthm}
\usepackage{latexsym}
\usepackage{color}
\usepackage{epsfig}
\usepackage{hyperref}
\usepackage{longtable}
\usepackage{multirow}
\usepackage{lscape}
\usepackage[normalem]{ulem}   % for use of \sout
\usepackage{ulem}
\usepackage[table]{xcolor}

\usepackage{Sweave}
\begin{document}

\Sconcordance{concordance:content.tex:C:/rufibach/20_Research/50_research_projects/130_estimands/10_FU_TTE/10_paper/content.Rnw:%
1 10 1 1 15 4 1 1 0 390 1 1 22 1 2 6 1 1 3 17 0 1 2 6 1 1 16 1 2 22 1 1 24 22 0 1 2 47 1 1 10 1 2 18 1 1 6 20 0 1 2 7 1 1 22 1 2 11 1 %
1 6 17 1 1 11 1 2 22 1 1 18 22 0 1 2 18 1 1 53 1 2 43 1 1 11 1 2 6 1 1 6 20 0 1 2 79 1}

\title{Quantification of follow-up time in oncology clinical trials with a time-to-event endpoint: Asking the right questions}

\author{
    Kaspar Rufibach\thanks{Methods, Collaboration, and Outreach Group (MCO), Product Development Data Sciences, Hoffmann-La Roche Ltd, Basel, Switzerland}\\
    Lynda Grinsted\thanks{AstraZeneca UK Ltd Cambridge, Cambridgeshire, UK}\\
    Jiang Li\thanks{BeiGene USA, Inc., 55 Challenger Road, Ridgefield Park, NJ 07660}\\
%    Yue Shentu\thanks{Daiichi Sankyo}\\
    Hans Jochen Weber\thanks{Clinical Development and Analytics, Novartis Pharma AG, Basel, Switzerland}\\
    Cheng Zheng\thanks{Zentalis Pharmaceuticals, New York, NY, USA}\\
    Jiangxiu Zhou\thanks{Statistics and Decision Sciences, J\&J, Spring House, PA, United States}
}

%\date{\today}

\maketitle

\begin{abstract}
For the analysis of a time-to-event endpoint in a single-arm or randomized clinical trial it is generally perceived that interpretation of a given estimate of the survival function, or the comparison between two groups, hinges on some quantification of the amount of follow-up. Typically, a median of some loosely defined quantity is reported. However, whatever median is reported, is typically not answering the question(s) trialists actually have in terms of follow-up quantification. In this paper, inspired by the estimand framework, we formulate a comprehensive list of relevant scientific questions that trialists have when reporting time-to-event data. We illustrate how these questions should be answered, and that reference to an unclearly defined follow-up quantity is not needed at all. In drug development, key decisions are made based on randomized controlled trials, and we therefore also discuss relevant scientific questions not only when looking at a time-to-event endpoint in one group, but also for comparisons. We find that different thinking about some of the relevant scientific questions around follow-up is required depending on whether a proportional hazards assumption can be made or other patterns of survival functions are anticipated, e.g. delayed separation, crossing survival functions, or the potential for cure. We conclude the paper with practical recommendations.
\end{abstract}


\textit{Running head:} Follow-up in time-to-event trials: Asking the right questions.

% % -----------------------------------------------------
% \section*{Administrative stuff}
% % -----------------------------------------------------
%
% \bi
% \item Targeted journal: Pharmaceutical Statistics? Statistical methods in clinical research? Trials? And key conclusions to JCO?
% \item Overleaf link: \url{https://www.overleaf.com/project/611a8f6012d753956721b459}.
% \item Markdown page with code: \url{https://numbersman77.github.io/quantFU/quantFU.html}.
% \item Github repository with code: \url{https://github.com/numbersman77/quantFU}.
% \ei
%
%
% \newpage
%
% \kr{TOC will be removed for submission, just here to see structure}
%
% \tableofcontents
%
% \newpage

% -----------------------------------------------------
\section{Introduction}
\label{intro}
% -----------------------------------------------------

For the analysis of a time-to-event (T2E) endpoint in a single-arm (SAT) or randomized controlled trial (RCT) it is generally perceived that interpretation of a given estimate of the survival function, or the comparison between two groups, hinges on some quantification of the amount of follow-up. This is particularly prevalent for oncology clinical trials, where T2E endpoints, such as progression-free (PFS) and overall survival (OS), are commonly used as primary endpoint. Positive results in such a trial typically lead to regulatory approval and, at least in many cases, to a change of standard clinical practice.

Often in reporting of such trials some quantification of follow-up is provided, typically via a median of some quantity. However, as Betensky\cite{betensky_15} reports, still recently in about half of reported trials in her review it is unclear what exactly this median is referring to and no interpretation beyond the reporting of this bare figure is given. As already discussed in Shuster\cite{shuster_91} whatever median is reported is typically not answering the question(s) trialists actually have in terms of follow-up quantification. While much has been said about this broad question in the literature (see Section~\ref{questions} for a review) in this paper we would like to add the following aspects:
\bi
% \item The estimand framework put forward in the ICH E9 estimand addendum\cite{ich_19} (referred to as "addendum" in the rest of the paper) has been broadly implemented in pharmaceutical drug development, so far primarily for efficacy endpoints. We have found that taking inspiration from the estimand framework to structure the question of follow-up for a T2E endpoint and all the quantities that have been proposed to "estimate" it allows for a transparent way of describing and appreciating the various approaches.
\item Following the thinking process in the ICH E9(R1) estimand addendum\cite{ich_19} we start with a comprehensive list of relevant {\it scientific questions} that trialists have when reporting T2E data and which are often "answered" with reference to some unclearly defined quantifier of follow-up. We illustrate how instead these questions should be answered, and that reference to an unclearly defined "follow-up quantity" is not necessary.
\item The literature so far has focused on quantifying follow-up for estimation of a survival function in one group. However, in oncology drug development key decisions are made based on RCTs, and we discuss relevant scientific questions in this context as well.
\item We define and illustrate a collection of follow-up quantifiers that are routinely used in reporting of trials. One of these, namely $\#7$ in Table~\ref{tab:quantities}, has not been discussed in the statistical literature previously.
\item Finally, with the advent of new therapeutic modalities especially in (but not restricted to) oncology, patterns of survival functions in trials emerged or became more prevalent, e.g. delayed separation, crossing survival functions, or the potential for cure, that may require different thinking on some of the relevant scientific questions.
\ei

While our focus will be on oncology, the findings are applicable in general to SATs or RCTs with a T2E endpoint in any therapeutic area.

In Section~\ref{questions} we draw on the literature and our experience as drug developers in oncology to list the questions that trialists typically would like to get answered with the broad concept of "follow-up". We then discuss in Section~\ref{existing} the various quantifiers of follow-up while Section~\ref{questions_really} makes the above loosely defined questions specific before we discuss two examples: one satisfying the proportional hazards (PH) assumption in Section~\ref{ex:gallium} and one with delayed separation in Section~\ref{ex:delayed}. We conclude with comments on the (ir)relevance of quantification of follow-up and general recommendations.

% -----------------------------------------------------
\section{What questions do trialists want to answer?}
\label{questions}
% -----------------------------------------------------

Let us assume we are interested in analyzing a T2E endpoint, such as PFS or OS, in a cohort of patients. Some patients had the event already, some patients may be lost to-follow-up (LTFU), and some are still under observation, i.e. remain administratively censored (= reporting event called by the sponsor). For sake of simplicity we assume that the survival function has been estimated using the Kaplan-Meier (KM) method, as this is by far the most used estimation approach for such data. Under appropriate assumptions on the censoring mechanism (e.g. random censoring) this is an unbiased estimate of the underlying true survival function, automatically accounting for varying follow-up times, as discussed e.g. by Pocock et al.\cite{pocock_02}

Now, a \km estimate of a survival function is typically accompanied by some quantification of "follow-up", where often neither the precise definition nor the interpretation of the latter is given\cite{betensky_15}. The first who looked into this question was Shuster \cite{shuster_91}. Based on interviews with oncologists he distilled the following questions that trialists are interested in when considering the role of median follow-up:

{\it
\begin{enumerate}
    \item "Maturity" of the estimated survival function.
    \item "Stability" of the estimated survival function.
    \item Time interval where \km estimate is "valid".
    \item "Quality" of follow-up.
\end{enumerate}
}

We put those parts of these questions in inverted commas that need a definition. In addition, the same oncologists were asked what "median follow-up" means for them, and the group came up with four different answers:

{\it
\begin{enumerate}
    \item Follow-up among those who did not have the event yet.
    \item Follow-up of all patients, irrespective of event status.
    \item Time from patient trial entry to clinical cut-off date.
    \item The censoring distribution, estimated through \km based on inverting the censoring indicator of the primary endpoint.
\end{enumerate}
}

However, as Shuster already concludes, the median of none of these quantities can be used to answer any of the above questions. Altman et al.\cite{altman_95_surv} and Schemper et al.\cite{schemper_96} performed systematic literature reviews where Altman et al. conclude that (1){\it "...almost half of the papers did not give any summary of length of follow-up"} and (2) {\it "For those which did give a summary, the median follow-up time was the most frequent value presented, although the method used to compute it was rarely specified."} The situation had not improved two decades later: Betensky\cite{betensky_15} repeated such a review and again found that {\it "34/60 trials did not say what they mean by median follow-up. All that reported anything only gave medians, none of the papers provided an interpretation."} Betensky concludes that {\it "At the very least, the measure that is reported must be clearly specified."}

The CONSORT statement\cite{moher_10} also has a short paragraph on follow-up, referencing Shuster \cite{shuster_91} and Altman et al.\cite{altman_95_surv}. They recommend for RCTs with a T2E endpoint to always give the date when follow-up ends, in addition to minimum, maximum, and median duration of follow-up, so basically a summary of the follow-up distribution. However, the statement is not specific as to {\it how} to compute follow-up.

We hope that this paper can bring some clarity and structure around relevant scientific questions, and how they can be answered.

We will now discuss what the literature offers as definition, if at all, for the four terms that had been quoted by Shuster. Recall that he got these as responses from trialists on the question about how they interpret "median follow-up".

% -----------------------------------------------------
\subsection{Maturity}
\label{maturity}
% -----------------------------------------------------

There is no uniformly accepted definition of what "maturity" means in this context. For example, Ben-Aharon et al.\cite{benaharon_20} seem to use "more mature" simply to mean "later snapshot" or "more data". On the other hand, Tai et al.\cite{tai_21} define maturity as the proportion of patients that had experienced an event, i.e. $\dint / n$ or $\dfin / n$, e.g. the proportion of patients who had died.
%This is also this group's understanding of "maturity". This, because if all patients had the event then we would have a sample of complete data (i.e. no censoring) which seems an appropriate concept of "full maturity".
However, this sometimes appears to be confused with the information fraction $\dint / \dfin$ at an interim analysis. As an example, an Oncologic Drugs Advisory Committee (ODAC) meeting of the US FDA in February 2021 assessed an RCT comparing pembrolizumab + chemotherapy vs. placebo + chemotherapy in high-risk early-stage triple-negative breast cancer (TNBC), the KEYNOTE-522\cite{schmid_20} trial. In the {\it Combined FDA and Applicant ODAC Briefing Document}\cite{odac_pembro} the trial data, based on an interim analysis for efficacy, was considered "immature" because only 174 / 327 = 53\% (for event-free survival) and 96 / 300 = 32\% (for OS) of the events planned for the final analysis had already been observed at this interim analysis. However, note that this trial recruited 1174 patients in total, so another way to look at "maturity" would be to take the latter number as denominator. It all basically boils down whether we want to assess "maturity" of estimation of the hazard ratio (HR, then information fraction appears appropriate) or the survival function in one or both groups. In the latter case, it is irrelevant which portion of targeted events for a hypothesis test for the HR we have already collected, but the proportion of patients that experienced an event matters more.

In any case, none of these quantities, i.e. neither $\dint / \dfin$ nor $\dint / n$ quantifies "follow-up" using some statistic (e.g. median) of a cumulative distribution function of a set of {\it times}, but rather based on {\it proportion of events} only. Also, as we discuss below, the actual "information" available at an interim analysis only exclusively depends on the number of events in case of PH. Things are different in case of non-PH.

"Maturity", understood as $\dint / n$ or $\dfin / n$, needs to be interpreted in connection with specifications of the underlying disease. For example, if we expect that 25\% of patients will be cured with the treatment under consideration then it is not reasonable to assume that the data from which this survival function is estimated can become more "mature" (as measured by some fraction of events on total number of patients) than 75\%.

If cure is possible in the disease under study the question of "enough follow-up" has been given a precise meaning in \cite{maller_94} and put in a hypothesis test. For a survival function $S$, let $\tau_S = \inf_{t \ge 0}\{S(t)  = 0\}$ (with the infimum of the empty set equal to $\infty$) denote the upper endpoint of its support. The null hypothesis considered in \cite{maller_94} is then
\bea
  H_0: \tau_{S_C} \ \le \ \tau_{S_X}.
\eea If $H_0$ is rejected we conclude that the support of the censoring distribution is not smaller than the one of the T2E endpoint and that we therefore have enough follow-up to detect a potential proportion of cured patients. Maller et al.\cite{maller_21} derive a general expression for the finite and large-sample distributions of a test statistic that has been proposed in Maller and Zhou\cite{maller_94}. However, practical experience first has to be gained with this approach, e.g. in terms of how powerful the test is in various scenarios. And the method addresses a quite specific scientific question, namely whether the survival function under consideration corresponds to a population with a proportion of cured individuals.

% -----------------------------------------------------
\subsection{Stability}
\label{stability}
% -----------------------------------------------------

Similar to "maturity" the term "stability" has been loosely (and interchangeably with the other terms) used in the literature until a formal definition was given by Betensky: "stability" is the relative information compared to the potential maximum information about the distribution of $X$ that is contained in the sample. Using the deterministic minimum and maximum of the \km estimate under assumed complete follow-up of censored observations, it quantifies the variability ({\it not} understood in a statistical sense, but rather how much the estimate can potentially change) of our current estimate relative to a future estimate given additional follow-up. Maximum information is achieved if all patients had the event, i.e. if we only have complete (as opposed to some of them being censored) observations. This can straightforwardly be quantified using the ratio of number of events to the number of patients. More specifically, "stability" may also refer to the question about how much an estimate of the survival function, typically \kmns, can still change until the next clinical cut-off date (CCOD) or the end of the trial. To this end, Betensky proposes to look at the most extreme scenarios assuming either all still censored patients will have an event on the next day or would be censored at the last observed event date. See Figure~\ref{fig:delayed_betensky} below for an example.

However, also "stability" is used to mean different things. As an example, Gebski et al.\cite{gebski_18} discuss methods to quantify "stability" (in Betensky's sense) but they call it "maturity".

% -----------------------------------------------------
\subsection{Validity}
\label{validity}
% -----------------------------------------------------

"Validity" is another vaguely or ambiguously defined concept. Intuitively, "validity" may be understood as "unbiased" and therefore be a general concept for any estimator. However, we read Definition 3 ("time interval of validity of the KM curve") in Shuster's letter such that his interpretation is that this refers to how {\it precise} a given estimate of the underlying survival function $S_X$ is. Again, as Shuster already emphasizes, this cannot really be answered by giving some statistics of some kind of "follow-up", however defined. Instead, as Betensky already discussed, the uncertainty of an estimate can be directly quantified using pointwise confidence intervals (CI) or confidence bands.

% -----------------------------------------------------
\subsection{Quality}
\label{quality}
% -----------------------------------------------------

Again, it is not clear what "quality" would refer to when it comes to "median follow-up". One could argue that "quality" refers to well-run trials, implying that only few patients are censored for other reasons than administrative censoring. This can be assessed by analyzing reasons for censoring, e.g. as discussed as Question 9 in Table~\ref{tab:examples2}. Again, if this were what "quality" referred to no median follow-up whatsoever can answer this question.

% -----------------------------------------------------
\section{Quantities used in the literature to quantify "follow-up"}
\label{existing}
% -----------------------------------------------------

Following Betensky\cite{betensky_15} let $X$ denote the time to event, $C$ the time to censoring, $T$ the observed time, that is, $T = \min\{X, C\}.$ The corresponding survival functions are $S_X(t) = P(X > t)$ and $S_C(t) = P(C > t)$. For the analysis of the primary endpoint of the trial we are interested in the survival function for $X$, $S_X$, which is typically estimated using the \km estimator.
% To this end, assume we have a sample of $n$ patients that provide data as follows:
% \bi
% \item $a_i$ the time when she was recruited,
% \item $b_i$ the observed event date for the patient, i.e. date when the patient was last confirmed to be censored, dropped from the trial, or exprienced the event of interest,
% \item $\delta_i = \begin{cases} 0 & \text{censored,} \\ 1 & \text{event of interest} \\ 2 & \text{LTFU}\end{cases}$
%
% is the censoring indicator for patient $i$.
% \ei
Furthermore, denote by $T_1, T_2$ the start and end of recruitment and by $T_3$ the CCOD, i.e. the date which the data will be reported up to. In the literature $T_3$ is typically referred to as "end of study" \cite{schemper_96}, \cite{betensky_15}, but we prefer the term CCOD. This, because a trial can have multiple reporting events, e.g. an efficacy interim and final analysis, where the former must not be the "end of study". Finally, assume that either in a SAT or RCT two CCODs are planned, both based on the primary efficacy endpoint:
\bi
\item An interim analysis after $\dint$ events.
\item A final analysis after $\dfin$ events.
\ei
Table~\ref{tab:quantities} summarizes the quantities that are used in the clinical literature to quantify "follow-up". Note that in the table a distinction is made for censored patients between the two clinical events "administrative censoring" (date where the patient was last confirmed event-free, e.g. at the last available event-free tumor assessment for PFS or date last known alive for OS) and "LTFU" or "drop-out", where a patient has left the trial prior to experiencing the event of interest and prior to $T_3$.

\begin{landscape}
\begin{table}[h]
\begin{center}
\begin{tabular}{|c|p{3.8cm}|*{3}{c|}p{2.4cm}|*{5}{c|}}\hline
\rule{0pt}{30pt} Number & Term                             & \parbox{1.4cm}{\centering Number in Shuster \cite{shuster_91}} & \parbox{1.65cm}{\centering Number in Schemper et al. \cite{schemper_96}} & \parbox{1.4cm}{\centering Betensky \cite{betensky_15}} & \parbox{1.7cm}{\centering Examples} & \parbox{1.2cm}{\centering Patient subset} & \parbox{1.4cm}{\centering Primary event} & \parbox{1.8cm}{\centering Censoring: administrative} & \parbox[c][2cm][c]{1.8cm}{\centering Censoring: LTFU} & \parbox{1.3cm}{\centering CCOD} \\ \hline
1 & Observation time regardless of censoring               & (2)               & 1. & $\min\{X, C\}$     & Example in Betensky \cite{betensky_15}&                & event             & event                         & event            & ignored \\ \hline
2 & \multirow{2}{*}{\parbox{3.8cm}{Observation time for those censored}} & (1) & 2. & $C|C<X$            & Example in Betensky \cite{betensky_15} & censored       & \cellcolor{gray}            & event                         & event            & ignored \\ \cline{7-11}
  &                                                        &                   &    &                    & & event          & \multicolumn{4}{c|}{excluded}                                       \\ \hline
3 & Time to censoring                                      & (4)               & 5. & $C$                & Gallium \cite{marcus_17} &                & censored             & event                         & event            & ignored \\ \hline
4 & Time to CCOD, Potential follow-up                      & (3)               & 3. &                    & Keynote-024 \cite{reck_16}*, Keynote-361\cite{powles_21} &                & ignored       & ignored                   & ignored      & event \\ \hline
5 & \multirow{2}{*}{\parbox{3.8cm}{Known function time}}   &                   & 4. &                    & None known. & censored       & \cellcolor{gray}             & event                         & event            & ignored \\ \cline{7-11}
  &                                                        &                   &    &                    & & event          & ignored       & \cellcolor{gray}                   & \cellcolor{gray}      & event       \\ \hline
6 & Korn's potential follow-up time                        &                   & 6. &                    & None known. &                & \multicolumn{4}{c|}{see below}                                     \\ \hline
%7 & Modified potential follow-up time                      &                   &    &                    & &                & ignored       & ignored                   & 1            & 1       \\ \hline
7 & \multirow{2}{*}{\parbox{3.8cm}{Potential follow-up considering events}}  &   &    &                    & Ascend-05 \cite{shaw_17}* & censored       & \cellcolor{gray}       & ignored                   & ignored      & event       \\ \cline{7-11}
 &                                                         &                   &    &                    & & event          & event             & \cellcolor{gray}                   & \cellcolor{gray}      & ignored \\ \hline
\end{tabular}
\caption{Quantities used in the literature to define "follow-up". Explanation of columns: \newline {\it Number in Shuster \cite{shuster_91} / Schemper et al. \cite{schemper_96}}: These are the numbers in these papers that were given to the respective quantities. Added here for easy identification. \newline {\it Patient subset}: Which patients to consider, based on their status for the primary endpoint. \newline The last four columns indicate how to treat these clinical events for the quantities of interest in each row. An entry of "1" means that this original datapoint is set to "event", "0" to censored, or "ignored". \newline {\it Censoring: administrative} refers to the date where the patient was last confirmed event-free, e.g. at the last available event-free tumor assessment for PFS or date last known alive for OS. "CCOD" refers to the clinical cut-off date.
\newline
* Definition of "follow-up" often not given in publicly available documents. In absence of publicly available details we know the definition either from personal communication of statisticians involved in these trials, or from comparing consecutive snapshots.}
\label{tab:quantities}
\end{center}
\end{table}
\end{landscape}

Typically, some quantile (primarily the median) of the distribution of the quantities in Table~\ref{tab:quantities} is reported in clinical publications\cite{schemper_96}. Looking at all the quantities and how differently they are defined it comes without a surprise that there is a lot of heterogeneity and confusion in the medical literature around follow-up, as exemplified in Sections~\ref{ex:gallium} and \ref{ex:delayed}.

Quantities \#1-\#6 have already been discussed by Schemper et al.\cite{schemper_96} and we refer to that paper for additional details.

Quantity \#4 is used in the Keynote-024 trial, with primary publication Reck et al.\cite{reck_16} and report of follow-up snapshot in Reck et al.\cite{reck_19}. We observe that the reported difference in median follow-up is {\it equal} to the difference between the two CCODs, which even applies to the ranges (i.e. minimum and maximum follow-up). This is virtually impossible to happen if other quantities were implemented than Quantity \#4. Although the actual calculation of follow-up time was not explicitly indicated in the publications, we consider it very likely that Quantity \#4 was used. Another example is Keynote-361 where Quantity \#4 is even explicitly defined is Powles et al.\cite{powles_21}: "...median follow-up, defined as time from randomisation to data cutoff."

Quantity \#6, Korn's potential follow-up time\cite{korn_86}, is a generalization of time to censoring (Quantity \#3) and estimates the probability to be under follow-up at a given time $t$. To this end, it makes a distinction between the two potential censoring reasons, namely LTFU and administrative censoring. While Korn's paper is regularly cited in the literature by papers which analyze RCTs, the citation is typically used to refer to time to censoring (see e.g. Blum et al.\cite{blum_19}). We are not aware of an actual computation of \#6 in a trial.

Quantity \#7 has, to the best of our knowledge, not been discussed in the statistical literature previously but is regularly used in oncology RCTs, e.g. in Ascend-05\cite{shaw_17}.

Another measure that has been proposed is Clarke's $C$ \cite{clark_02} and variants thereof \cite{wu_08}. The former is simply the ratio $\# 1 / \#7$ of the quantities in Table~\ref{tab:quantities} and the latter attenuates the {\it Potential follow-up considering events} in the denominator, Quantity $\#7$, to account for unobserved deaths.

While the primary reason to look at some quantification of follow-up is to understand features of survival functions or effect estimates it is still worthwhile mentioning that only two of the quantities in Table~\ref{tab:quantities} actually estimate a well-defined population quantity: Time to censoring \#3, which estimates $S_C$, the hypothetical follow-up at any time $t$ if no events would happen. And Korn's potential follow-up \#6 which estimates the probability to be under follow-up at $t$. All the other quantities either condition on censoring/event status and/or depend on the CCOD and as such have to be considered merely descriptive.

Based on our experience as statisticians in pharmaceutical drug development, quantities \#1 and \#3 are those that are most often used to report "follow-up". We have added the other quantities to the table because they had already been discussed by Shuster\cite{shuster_91}, Schemper et al.\cite{schemper_96}, and Betensky\cite{betensky_15}, or are being used in clinical trials.

To conclude this section we would like to share the observation that the approach by which "follow-up" is computed is typically not explicitly described in primary publications or trial protocols. However, these are the only documents that are often publicly available (e.g. for the Gallium trial: primary publication \cite{marcus_17} and protocol in accompanying supplementary material \cite{marcus_17_supp}). This makes it very difficult to find examples for each approach outlined in Table~\ref{tab:quantities}.

% -----------------------------------------------------
\section{Recommendations how to answer questions about maturity, stability, validity, and quality}
\label{questions_really}
% -----------------------------------------------------

The ICH E9(R1) estimand addendum\cite{ich_19} defines its applicability as
\begin{quote}
{\it The principles outlined in this addendum are relevant whenever a treatment effect is estimated, or a hypothesis related to a treatment effect is tested, whether related to efficacy or safety.}
\end{quote}
So in theory a direct application of the ICH E9(R1) estimand addendum\cite{ich_19} to the problem of quantifying follow-up does not appear indicated because we are not estimating or testing a treatment effect. But we still found the estimand thinking process useful for this problem, for two reasons:

\begin{enumerate}
    \item The emphasis on the {\it scientific objective} - what question do we want to answer for our primary T2E endpoint? As we describe below many implicit questions have been "answered" using some "median follow-up", without properly connecting the analysis to the question.
    \item The concept of Table~\ref{tab:quantities} is clearly inspired by the handling of {\it intercurrent events} when defining an estimand. Granted, not all of the clinical events (primary endpoint event, LTFU, administrative censoring, CCOD) we consider in Table~\ref{tab:quantities} are intercurrent events in the spirit of the addendum, but the thinking process and structure of the addendum remains applicable.
\end{enumerate}

In Sections~\ref{questions1} and \ref{questions2} we list potential questions of interest that trialists have when they talk about "follow-up". With "milestone" we refer to a timepoint of interest.

% -----------------------------------------------------
\subsection{One-sample case}
\label{questions1}
% -----------------------------------------------------

Having tried to give meaning to what questions trialists have for "median follow-up" as reported by Shuster, Table~\ref{tab:examples1} summarizes the questions one can ask when considering the \km estimate of a survival function in one group. "Precision" and "stability" had been discussed in detail in Betensky\cite{betensky_15} already.

\begin{table}[h]
\begin{center}
\begin{tabular}{c|l|p{5cm}|p{5cm}}\hline
Number & Term        & Question & How to best answer \\ \hline
1      & Precision   & How precise is an estimate with respect to the underlying true survival function $S_X$? & Pointwise CIs or confidence bands. \\ \hline
2      & Reliability & How far out to extend the \km estimate, as discussed in Pocock et al. \cite{pocock_02}.  & Pointwise CIs or confidence bands. Disregard \km estimate if less than $m$ patients remain at risk.\\ \hline
3      & Stability   & How much can a \km estimate possibly change in a future data snapshot?         & Consider all currently censored patients to either (1) have an event the day after censoring or (2) being censored at the latest observed event time. \\ \hline
4      & Information & How much of the information necessary to achieve a targeted power for a hypothesis test, either for a milestone timepoint or the median, has been collected? & Power depends on various quantities, see below. \\ \hline
\end{tabular}
\caption{Questions trialists want to answer for a \km estimate of a survival function in one sample, and how they are best answered.}
\label{tab:examples1}
\end{center}
\end{table}

Table~\ref{tab:examples1} makes it clear that none of the quantities described by Shuster and listed in Section~\ref{questions} is needed or useful to answer Questions 1 - 4. As for Question 2 we find that "reliability" best describes what Pocock et al.\cite{pocock_02} were after. They recommended "that survival plots be halted once the proportion of patients free of an event, but still in follow-up, becomes unduly small." and "It will often be reasonable to curtail the plot when only around 10-20\% are still in follow-up." and already mention that their view might not be universally held. Indeed we recommend that instead of curtailing the plot a proper quantification of uncertainty through pointwise CIs or a confidence band for the entire estimated function should be preferred. In our view, there is a big difference in terms of what having 20\% of patients still in follow-up in a trial with 1000 or 100 patients tells us about the underlying true survival function $S_X$. The decreased uncertainty in the larger trial is then properly reflected in the CIs or bands. Taking this even further, a recommendation for practitioners could be to disregard any \km estimate if less than some number of patients remain at risk. This number can e.g. be derived from how wide a confidence interval trialists are still willing to accept. That would make the consideration independent of the total number of patients in the trial and would allow determination of when to start ignoring the curve solely based on the number of patients still at risk, a quantity that is often accompanying \km estimates as in our Figures~\ref{fig:gallium} and \ref{fig:ex2}.

Hypothesis tests for Question 4 have been described in the literature, e.g. in Brookmeyer and Crowley\cite{brookmeyer_82} for the median (construct test from confidence interval) and Nagashima et al.\cite{nagashima_21} for a milestone timepoint. Such tests depend on quantities other than only the number events and we refer to these publications as well as the discussion for restricted-mean survival time (RMST) in Section~\ref{q2_noph} for details.

% -----------------------------------------------------
\subsection{Two-sample case}
\label{questions2}
% -----------------------------------------------------

For the two-sample case the primary question is not about estimates of survival functions anymore but rather about the comparison of survival functions to quantify a (precisely defined) treatment effect. Stability remains an issue as well for the comparison of two groups, as discussed by Betensky\cite{betensky_15}. When it comes to Betensky's definition of stability it may be useful to base this assessment on the proposed upper and lower limits for the individual \km estimates. We also found it useful to make a distinction between the PH and non-PH scenario for precision and reliability.

% -----------------------------------------------------
\subsubsection{Proportional hazards}
\label{q2_ph}
% -----------------------------------------------------

We start with making the PH assumption for the underlying hazard functions and use the HR as quantifier in the corresponding columns in Table~\ref{tab:examples2}. Typically, sample size computation is performed using the formula proposed by Schoenfeld\cite{schoenfeld_83}. We note that accuracy of this formula depends on
\bi
\item the PH assumption,
\item use of unweighted logrank test,
\item and the randomization ratio.
\ei
Alternative normal approximations of the distribution of the test statistic if one or more of these assumptions are not met are given in Yung and Liu\cite{yung_20}. However, for the purpose of this paper, we make the first two assumptions and assume a fixed randomization ratio. It is important to note that then powering a trial exclusively based on number of events is tightly connected to using the HR as an effect measure. Because only for the HR and assuming PH the precision of the effect quantifier {\it exclusively} depends on the number of events (and via a constant on the randomization ratio). This then allows to set up a group-sequential design using the number of events as the "metric of information", because the statistical information is defined as the inverse of the variance of the group difference parameter, see e.g. Lan and Zucker\cite{lan_93}. So, as already discussed in Section~\ref{maturity}, the information fraction $\dint / \dfin$ then quantifies how much of the targeted events that were pre-specified in the sample size computation we have already reached, e.g. at an interim analysis. We reiterate that "information fraction" is not to be confused with "follow-up". This directly leads to Question 7 in Table~\ref{tab:examples2}.

% This is the old Table 3 that has been merged with the old Table 3 in the 2nd review round:
% \begin{table}[h]
% \begin{center}
% \begin{tabular}{c|l|p{5cm}|p{5cm}}\hline
% Number & Term        & Question & How to best answer \\ \hline
% 5a      & Precision   & How precise is a HR estimate? & CI. \\ \hline
% 6a      & Stability   & How much can an estimate of a HR change in a future data snapshot? & If the PH assumption applies then an estimate of the HR will (on average) simply become more precise over time. \\ \hline
% 7a      & Information & How much of the information necessary to achieve a targeted power for a hypothesis test for the HR within a group-sequential design has already been collected? & Information fraction $\dint / \dfin$.\\ \hline
% 8a      & PH (reliability)          & Do hazard functions remain proportional? & Standard tools to assess PH, e.g. plot nonparametric estimates of (cumulative) hazard functions over time, and ratio thereof, or hypothesis tests. \\ \hline
% 9a      & Censoring pattern          & Is censoring distribution the same in the two arms? Are distributions of censoring reasons the same in the two arms? & Plot nonparametric estimates of the censoring distribution $S_C$ per arm, potentially split by censoring reason.  \\ \hline
% \end{tabular}
% \caption{PH scenario: Questions trialists want to answer for the primary T2E endpoint, and how they are best answered.}
% \label{tab:examples2}
% \end{center}
% \end{table}

\newpage

\begin{landscape}
\begin{table}[h]
\centering
\begin{tabular}{|c|p{2cm}|p{4.5cm}|p{4.5cm}|p{4.5cm}|p{4.5cm}|}\hline
\multirow{2}{1.3cm}{Number} & \multirow{2}{1.5cm}{\hspace*{0.3cm}Term}        & \multicolumn{2}{c|}{Question}     & \multicolumn{2}{c|}{How to best answer}      \\ \cline{3-6}
                        &             & Proportional hazards                     &  Non-prop. hazards & Proportional hazards               & Non-prop. hazards \\ \hline
5                      & Precision   & How precise is estimate of HR? & How precise is chosen estimate to quantify effect of interest?        & \multicolumn{2}{c|}{Confidence interval.} \\ \hline
6                      & Stability   & How much can an estimate of a HR change in a future data snapshot? & How much can an estimate of the effect of interest change in a future data snapshot? & If the PH assumption applies then an estimate of the HR will (on average) simply become more precise over time. & Look at extreme scenarios as proposed by Betensky\cite{betensky_15}. \\ \hline
7      & Information & \multicolumn{2}{p{9cm}|}{How much of the information necessary to achieve a targeted power for a hypothesis test for the effect of interest (HR if PH) has already been collected, if a group-sequential design is used?}  & Information fraction $\dint / \dfin$. & Not only related to the information fraction, effect measure specific. \\ \hline
8      & Reliability          & Do hazard functions remain proportional? & Not applicable. & Standard tools to assess PH, e.g. plot nonparametric estimates of (cumulative) hazard functions over time, and ratio thereof, or hypothesis tests. &\\ \hline
9      & Censoring pattern          & \multicolumn{2}{p{9cm}|}{Is censoring distribution the same in the two arms? Are distributions of censoring reasons the same in the two arms?} & \multicolumn{2}{p{9cm}|}{Plot nonparametric estimates of the censoring distribution $S_C$ per arm, potentially split by censoring reason.}\\ \hline
\end{tabular}
\caption{Questions trialists want to answer for the primary T2E endpoint, and how they are best answered. For PH and NPH scenario.}
\label{tab:examples2}
\end{table}
\end{landscape}

\newpage

As for Question 6 in Table~\ref{tab:examples2} one could argue that what we are suggesting as the best answer for PH is not really addressing stability, as it does not consider the most extreme possibilities as Question 2 in Table~\ref{tab:examples1}. However, it is important to realize that if we entertain the PH assumption over the entire observation time then we restrict the range within which an early estimate of the HR can vary considerably. Additional follow-up will only give us a more precise estimate, and it is unlikely that this early HR estimate will dramatically change. This implies that if we have a trial that stops "early" in the sense that only a low proportion of patients already had their event, stability is more driven by our {\it belief} in the PH assumption rather than in what can possibly happen with more follow-up. This is especially relevant if a trial stops at an interim analysis for efficacy. Question 9 may shed some light on whether informative censoring might be present, especially in open-label trials. All these aspects will be discussed further using a real trial example in Section~\ref{ex:gallium}.

We are well aware that the above interpretation, exclusively relying on the PH property to "predict" the future course of the trial, is a strong and statistically motivated assumption. From a broader drug development perspective (or from the perspective of a Health Authority) it is acknowledged that after a first CCOD the stability of a potential treatment effect needs to be assessed also based on accumulating data, not "only" on the PH assumption. However, the point is that such stability cannot be assessed using whatever measure of follow-up. The best way is to look at nonparametric estimates of survival functions (i.e \kmns).

% -----------------------------------------------------
\subsubsection{Non-proportional hazards}
\label{q2_noph}
% -----------------------------------------------------

Now, if hazards are not assumed to be proportional, trialists have to quantify the effect using a measure other than, or in addition to, the HR, depending on the anticipated nature of the PH violation. The most prominent approaches are either a comparison (through the difference or ratio) of survival functions at a given milestone timepoint, e.g. one year, a comparison of medians, the use of RMST, i.e. the area between two estimates of the survival functions, or some weighted version of the logrank test (to which we also add max-combo tests). See e.g. Uno et al.\cite{uno_14} for an introduction into the first two measures, for max-combo tests Roychoudhury et al.\cite{roychoudhury_21}, Yung and Liu\cite{yung_20} for a discussion of statistical properties of these test statistics, and Lin et al.\cite{lin_20} for a comparison of the properties of several methods. Typically, RMST is computed based on \km estimates and is therefore truly nonparametric, i.e. does not make an assumption about the treatment group having a hazard function proportional to the one in the control group. However, as the name suggests, it depends on choosing (either a priori or data-driven) a timepoint $t_0$ that restricts the observation time over which the area between the two estimates is compared, so is more explicitly connected to the amount of follow-up already. But as mentioned by A'Hern\cite{ahern_16}, the use of RMST does not overcome at all any of the problems posed by limited patient follow-up.

For both RMST and the difference or ratio of $t$-year event-free probabilities, which we will call {\it milestone probabilities}, precision of these estimates is again best quantified using CIs for these quantities. For the latter it is important, as discussed by Pocock et al.\cite{pocock_02}, to consider CIs for the {\it difference} of the two estimated event-free probabilities, and not only provide CIs around each individual estimated survival function at $t$.

As we have elaborated in Section~\ref{q2_ph} the PH assumption allows a group-sequential design to be run exclusively based on number of events as the basic measure of "information". For other measures of effect this is not necessarily the case: information then needs to be more generally considered as the inverse of the variance of the group difference parameter. As an example, Hasegawa et al.\cite{hasegawa_20} discuss this for RMST at a restriction timepoint $t_0$ and provide a formula for the information initially derived by Murray and Tsiatis\cite{murray_99}. This expression for the information is much more complicated than for the HR and depends on {\it all} of the following quantities:
\bi
\item total number of patients,
\item randomization ratio,
\item \km estimate of the pooled sample,
\item estimated censoring distribution in each arm (which can be taken as pooled if random censoring is assumed),
\item observed number of events at $t_0$,
\item observed number of patients still at risk at $t_0$.
\ei
That for non-PH the power of a hypothesis test does not anymore exclusively depend on the number of events, but also on the accrual and censoring pattern, has already been discussed by Struthers and Kalbfleisch\cite{struthers_86}. So there is no hope that one can provide an accurate summary of "information" when running a group-sequential design for RMST with whatever quantification of follow-up. Also, it is very important to recognize that this implies that for non-PH scenarios, number of events will typically {\it not} be a sufficient metric to quantify information.

Lu and Tian\cite{lu_21} discuss how to plan and run a group-sequential design for RMST while Wang et al.\cite{wang_21} discuss a simulation-free group-sequential design with max-combo tests in the presence of non-proportional hazards.

A common pattern for non-PH is in immuno-oncology with {\it delayed separation}, see e.g. Mick and Chen\cite{mick_15}, i.e. survival functions only start to separate after a certain time. Potentially, we may also expect or observe crossing survival functions or that patients can be cured, see Section~\ref{maturity}. Collection of hypothesis tests for crossing survival functions are discussed in Li et al.\cite{li_15_crossing} and Dormuth et al.\cite{dormuth_22}. With all these features the first consideration for trialists is to decide on a quantity with which they want to quantify a potential treatment effect and perform a corresponding hypothesis test. For all of them it is particularly important to be able to characterize the tail behavior of survival functions out to some milestone timepoint that is clinically relevant. One needs to ensure enough follow-up such that the chosen hypothesis test achieves sufficient power, be it for RMST, milestone survival, a weighted logrank, or max-combo test. We summarize the questions and recommendations for the non-PH scenario in Table~\ref{tab:examples2}.

% This is the old Table 4 that has been merged with the old Table 3 in the 2nd review round:
% \begin{table}[h]
% \begin{center}
% \begin{tabular}{c|l|p{5cm}|p{5cm}}\hline
% Number & Term        & Question & How to best answer \\ \hline
% 5b      & Precision   & How precise is the chosen estimate to quantify the effect of interest? & CI. \\ \hline
% 6b      & Stability   & How much can an estimate of the effect of interest change in a future data snapshot? & Look at extreme scenarios as proposed by Betensky\cite{betensky_15}. \\ \hline
% 7b      & Information & How much of the information necessary to achieve a targeted power for a hypothesis test for the effect of interest has already been collected, if a group-sequential design is used? & Not only related to the information fraction, effect measure specific.\\ \hline
% 8b      & PH          & Not applicable. \\ \hline
% 9b      & Censoring pattern          & Same as in PH scenario. \\ \hline
% \end{tabular}
% \caption{Non-PH scenario: Questions trialists want to answer for the primary T2E endpoint, and how they are best answered.}
% \label{tab:examples2nph}
% \end{center}
% \end{table}

% -----------------------------------------------------
\section{Example with proportional hazards: the Gallium trial}
\label{ex:gallium}
% -----------------------------------------------------

We will now illustrate all these quantities using two clinical trial examples, one with approximately PH and one with non-PH.

\begin{figure}[h!tb]
\begin{center}
\setkeys{Gin}{width=1\textwidth}
\includegraphics{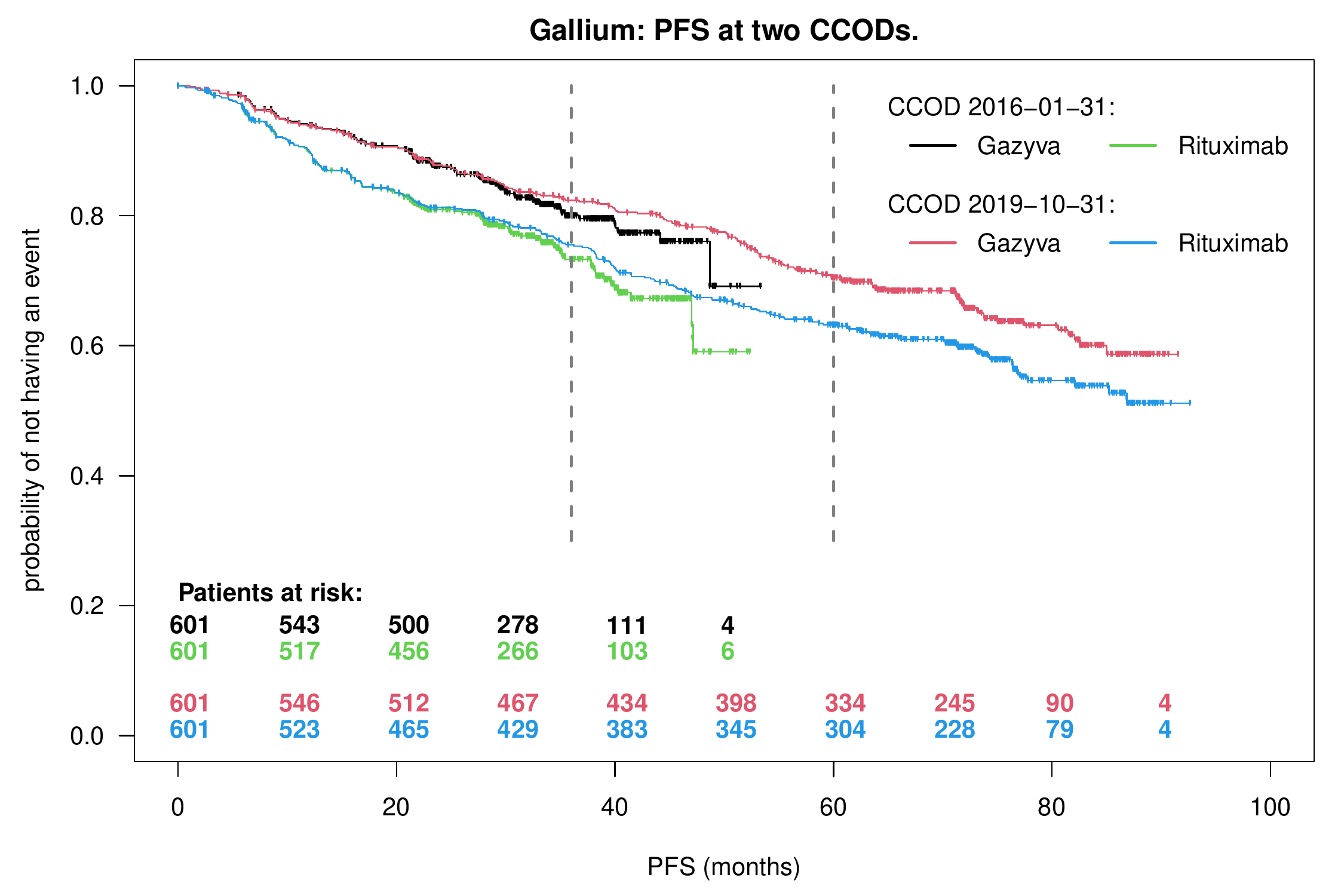}
\caption{PFS for Gallium. Vertical lines at milestone timepoints 36 and 60 months.}
\label{fig:gallium}
\end{center}
\end{figure}

\begin{figure}[h!tb]
\begin{center}
\setkeys{Gin}{width=1\textwidth}
\includegraphics{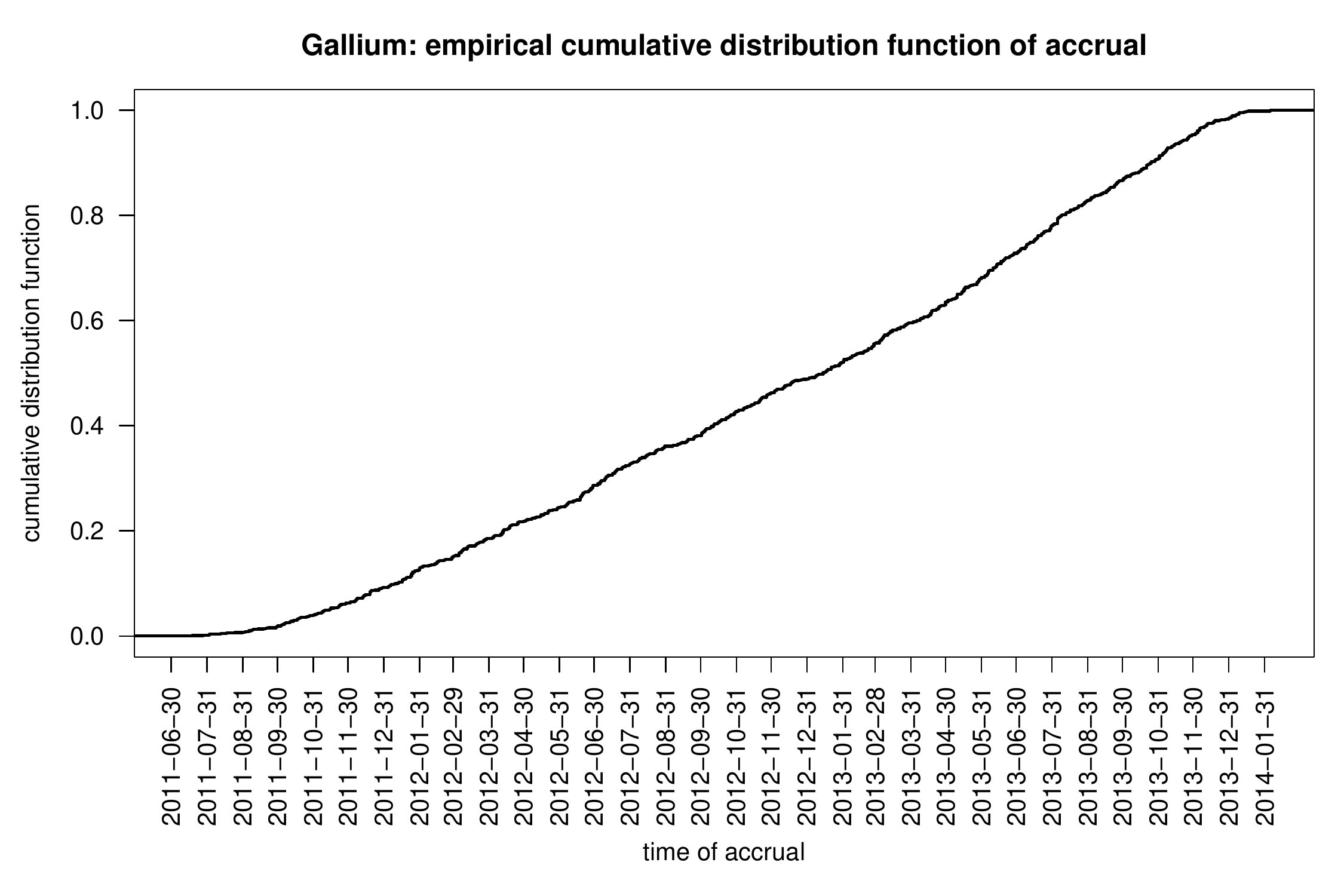}
\caption{Empirical cumulative distribution function of accrual for Gallium.}
\label{fig:gallium_accrual}
\end{center}
\end{figure}

We use the Gallium trial\cite{marcus_17} to exemplify the situation for (initially) PH. Gallium was a Phase 3, 1:1 RCT in first-line follicular lymphoma, comparing Rituximab (= control) to Obinutuzumab (= experimental) using a primary endpoint of PFS. Since administration schedule for the two drugs was different the trial was open-label for patients and physicians, but not for the sponsor. The trial was fully unblinded after the hypothesis test for the HR was statistically significant within a group-sequential design at a pre-planned interim analysis for efficacy. This interim analysis happened after 245 / 1202 (= 20.4\% of patients, information fraction = 245 / 370 = 0.66) patients had experienced a PFS event. We compare this snapshot to an update that was reported almost four years later\cite{townsend_20}. \km estimates of both snapshots are provided in Figure~\ref{fig:gallium}. For both snapshots Obinutuzumab showed a benefit compared to Rituximab. Key efficacy results of the trial are given in Table~\ref{tab:gallium efficacy}. Figure~\ref{fig:gallium_accrual} provides the empirical cumulative distribution function of accrual. This may be useful to check any unusual patterns in recruitment to the trial.

\begin{center}
\renewcommand{\baselinestretch}{1.1}
% latex table generated in R 4.2.2 by xtable 1.8-4 package
% Mon Mar 13 23:13:01 2023
\begin{table}[h]
\centering
\begingroup\normalsize
\begin{tabular}{lcc}
  \hline
 & CCOD 2016-01-31 & CCOD 2019-10-31 \\
  \hline
HR & 0.66 & 0.76 \\
  95$\%$ CI & [0.51, 0.85] & [0.62, 0.92] \\
  Number of events $d$ & 245 & 419 \\
  Proportion of patients with event & 20.4$\%$ & 34.9$\%$ \\
   \hline
\end{tabular}
\endgroup
\caption{Key efficacy results for Gallium.}
\label{tab:gallium efficacy}
\end{table}\renewcommand{\baselinestretch}{1}
\end{center}

% -----------------------------------------------------
\subsection{Answering the scientific questions}
\label{gallium:q}
% -----------------------------------------------------

% -----------------------------------------------------
\paragraph{Question 5: Precision.}
\label{gallium:q5}
% -----------------------------------------------------

Results for the HR are given in Table~\ref{tab:gallium efficacy}. Clinically, estimates of event-free probabilities at certain milestones are also of interest, e.g. at 36 and 60 months, respectively. We provide these, including their differences and 95\% CIs in Table~\ref{tab:gallium_milestones}. With these everything that might be of interest in terms of precision for these estimates is said.

As discussed in Section~\ref{questions1} we might also simply disregard \km estimates once less than a given number remain at risk in each arm, where this number can e.g. be informed by what confidence interval width trialists are willing to accept. This would entail that for the \km estimates in Figure~\ref{fig:gallium} there is no reason to worry about the drops very late in the estimates for the earlier CCOD. And indeed, with more data, i.e. at the later CCOD, these drops have been "overwhelmed" and disappeared.

\begin{center}
\renewcommand{\baselinestretch}{1.1}
% latex table generated in R 4.2.2 by xtable 1.8-4 package
% Mon Mar 13 23:13:01 2023
\begin{table}[h]
\centering
\begingroup\normalsize
\begin{tabular}{cccc}
  \hline
milestone & treatment arm & CCOD 2016-01-31 & CCOD 2019-10-31 \\
  \hline
 & Rituximab & 0.73 [0.66, 0.80] & 0.76 [0.71, 0.79] \\
  36
months & Obinutuzumab & 0.80 [0.73, 0.85] & 0.82 [0.79, 0.86] \\
   & Difference Obinutuzumab - Rituximab & 0.07 [0.01, 0.12] & 0.07 [0.02, 0.12] \\
   \hline
 & Rituximab & - & 0.63 [0.58, 0.68] \\
  60
months & Obinutuzumab & - & 0.70 [0.65, 0.75] \\
   & Difference Obinutuzumab - Rituximab & - & 0.07 [0.02, 0.13] \\
   \hline
\end{tabular}
\endgroup
\caption{Milestone KM estimates for Gallium.}
\label{tab:gallium_milestones}
\end{table}\renewcommand{\baselinestretch}{1}
\end{center}

% -----------------------------------------------------
\paragraph{Question 6: Stability.}
\label{gallium:q6}
% -----------------------------------------------------

If one believes in the PH assumption then in theory additional follow-up will not relevantly change the estimate of the HR but only make the CI more narrow. For the first Gallium snapshot, a hypothesis test for the null hypothesis of PH based on the scaled Schoenfeld residuals gives a $p$-value of 0.35, so little evidence for a deviation from PH.

Townsend et al.\cite{townsend_20} reported updated results based on 419 PFS events. The HR estimate now was 0.76 with 95\% CI [0.62, 0.92], see Table~\ref{tab:gallium efficacy}. So indeed, the CI is now more narrow, as a consequence of more information. Of note, the fact that the trial was fully unblinded and superiority of Obinutuzumab was claimed after the first CCOD may introduce bias for the estimation of the PFS effect of the initially assigned treatment. This e.g. through different patterns of use of new anti-lymphoma therapy (NALT) prior to progression (in the language of the ICH E9 estimands addendum\cite{ich_19} a treatment policy strategy was used for the intercurrent event of NALT). Thus, whether the increase of the HR estimate from 0.66 to 0.76 reflects a real change in effect potentially related to the potential bias introduced through unblinding or is simply due to variability in estimation cannot be decided based on the PFS data only. If we {\it believe} in the PH assumption and assume no bias then it would be variability only.

For the updated snapshot, the test for PH gives now a $p$-value of 0.03, so according to Pocock et al.\cite{pocock_15_1} this is "some evidence" against PH.

However, no quantification of follow-up whatsoever can (1) provide information on stability, (2) help decide the question about whether bias has been introduced after the first CCOD and (3) whether the PH assumption holds up to the second CCOD.

We also are of the opinion that if a PH assumption is entertained it is not informative to look at Betensky's stability criterion, as the extreme scenarios proposed there do not really allow effect quantification via HR. These criterion may be useful in the case of non-PH though, see Section~\ref{delayed:q6}.

% -----------------------------------------------------
\paragraph{Question 7: Information.}
\label{gallium:q7}
% -----------------------------------------------------

As discussed in Section~\ref{q2_ph} "information" or "information fraction" generally refers to the inverse of the variance of the group difference parameter in a group-sequential design. For PH this simplifies to the proportion of events that have been collected with respect to the total number of events needed to achieve the targeted power. In Gallium 370 events were needed to achieve 80\% power to detect a HR of 0.741 based on a two-sided logrank test with significance level of 0.05 and an O'Brien-Fleming group-sequential boundary for one efficacy interim analysis after about 2/3 of information. The efficacy interim analysis was planned to happen after 248 PFS events and actually took place after 245 events, so the information fraction at the interim analysis was 245 / 370 = 0.66.

The second snapshot was taken to update estimates of survival functions and the HR. Since the trial had rejected the null hypothesis of no difference between the arms already at the interim analysis this updated analysis happened outside of the group-sequential design and it is therefore meaningless to talk about "information fraction" in this context.

Again, also here, these considerations are completely independent of any quantification of follow-up whatsoever.

% -----------------------------------------------------
\paragraph{Question 8: Reliability.}
\label{gallium:q8}
% -----------------------------------------------------

This aspect has already been discussed in connection with "stability" in Section~\ref{gallium:q6}.

% -----------------------------------------------------
\paragraph{Question 9: Comparison of censoring distributions.}
\label{gallium:q8}
% -----------------------------------------------------

\begin{figure}[h!tb]
\begin{center}
\setkeys{Gin}{width=1\textwidth}
\includegraphics{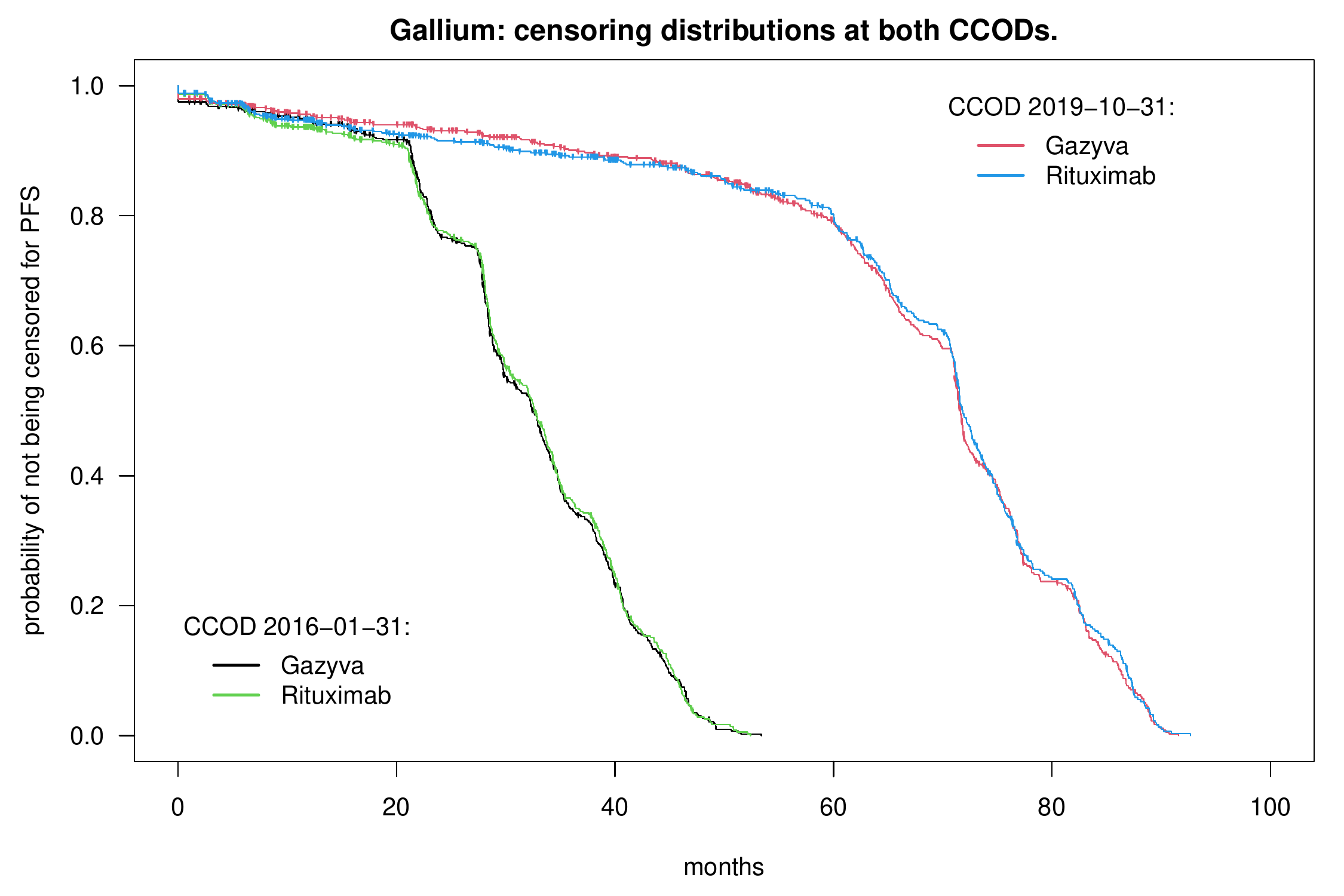}
\caption{Estimates of censoring distributions $S_C$ for Gallium.}
\label{fig:gallium_cens_arm}
\end{center}
\end{figure}

Finally, Figure~\ref{fig:gallium_cens_arm} compares censoring distributions between arms, estimated using the reverse \km method. We conclude that the censoring pattern was virtually identical between the two arms at both CCODs.

% -----------------------------------------------------
\subsection{Quantification of follow-up}
\label{gallium:qfu}
% -----------------------------------------------------

For completeness, and not because we think they provide any additional information with respect to the scientific questions in Section~\ref{gallium:q}, we provide all the follow-up quantifiers from Table~\ref{tab:quantities} in Table~\ref{tab:gallium}. Note that only Quantity \#6 (Korn's potential follow-up) in Table~\ref{tab:quantities} makes a distinction between {\it administrative censoring} and {\it LTFU}. However, for the Gallium data we did not have this information available in our dataset on a patient level, but from the CONSORT diagram in Figure~1 in Hiddemann et al.\cite{hiddemann_18} we find that only a very small portion of patients (derived from the reasons given for withdrawal) has to be considered LTFU. In our example, for illustrative purposes, we therefore randomly assigned the status of "LTFU" to 5\% of those patients in the trial who were administratively censored (separately at each CCOD). From Table~\ref{tab:gallium} we then make the following observations:
\bi
\item Although all the quantities in Table~\ref{tab:quantities} quantify some sort of follow-up there is relevant heterogeneity among them, with values ranging from 28.8 to 36.5 months and 62.0 to 81.5 months for the two CCODs, respectively. This spread matches the findings in the example discussed in Betensky\cite{betensky_15} and in the simulation study by Schemper et al.\cite{schemper_96}.
\item Not only are the actual values different, but also the increase from one snapshot to the next, either quantified as difference in months (column $\Delta$) or as relative increase (column $\Delta \%$). The former varies from 33.2 to 45 months and the latter from 115$\%$ to 128$\%$.
\item The difference between the two CCODs in calendar time is 45.0 months. However, that number is not directly connected to any of the follow-up quantifiers or their increase in Table~\ref{tab:gallium}, except to the increase for {\it Time to CCOD}, this is of course exactly 45.0 months by definition as long as the accrual was finished at the first CCOD (which was the case for Gallium). Rather, the increase in follow-up primarily depends on the frequency of, in the case of PFS, tumor assessments or confirmation of survival status for OS.
\ei

\begin{center}
\renewcommand{\baselinestretch}{1.1}
% latex table generated in R 4.2.2 by xtable 1.8-4 package
% Mon Mar 13 23:13:01 2023
\begin{table}[h]
\centering
\begingroup\normalsize
\begin{tabular}{cp{5.5cm}p{2cm}p{2cm}cc}
  \hline
Number & Quantity & CCOD 2016-01-31 & CCOD 2019-10-31 & $\Delta$ & $\Delta \%$ \\
  \hline
1 & Observation time regardless of censoring & 28.8 & 62.0 & 33.2 & +115$\%$ \\
  2 & Observation time for those event-free & 31.5 & 71.2 & 39.8 & +126$\%$ \\
  3 & Time to censoring & 32.6 & 71.7 & 39.1 & +120$\%$ \\
  4 & Time to CCOD & 36.5 & 81.5 & 45.0 & +123$\%$ \\
  5 & Known function time & 32.9 & 75.0 & 42.0 & +128$\%$ \\
  6 & Korn potential follow-up & 36.2 & 81.2 & 44.9 & +124$\%$ \\
  7 & Potential follow-up considering events & 33.4 & 75.5 & 42.1 & +126$\%$ \\
   \hline
\end{tabular}
\endgroup
\caption{Different quantifications of follow-up for Gallium, in months.}
\label{tab:gallium}
\end{table}\renewcommand{\baselinestretch}{1}
\end{center}

\begin{figure}[h!tb]
\begin{center}
\setkeys{Gin}{width=1\textwidth}
\includegraphics{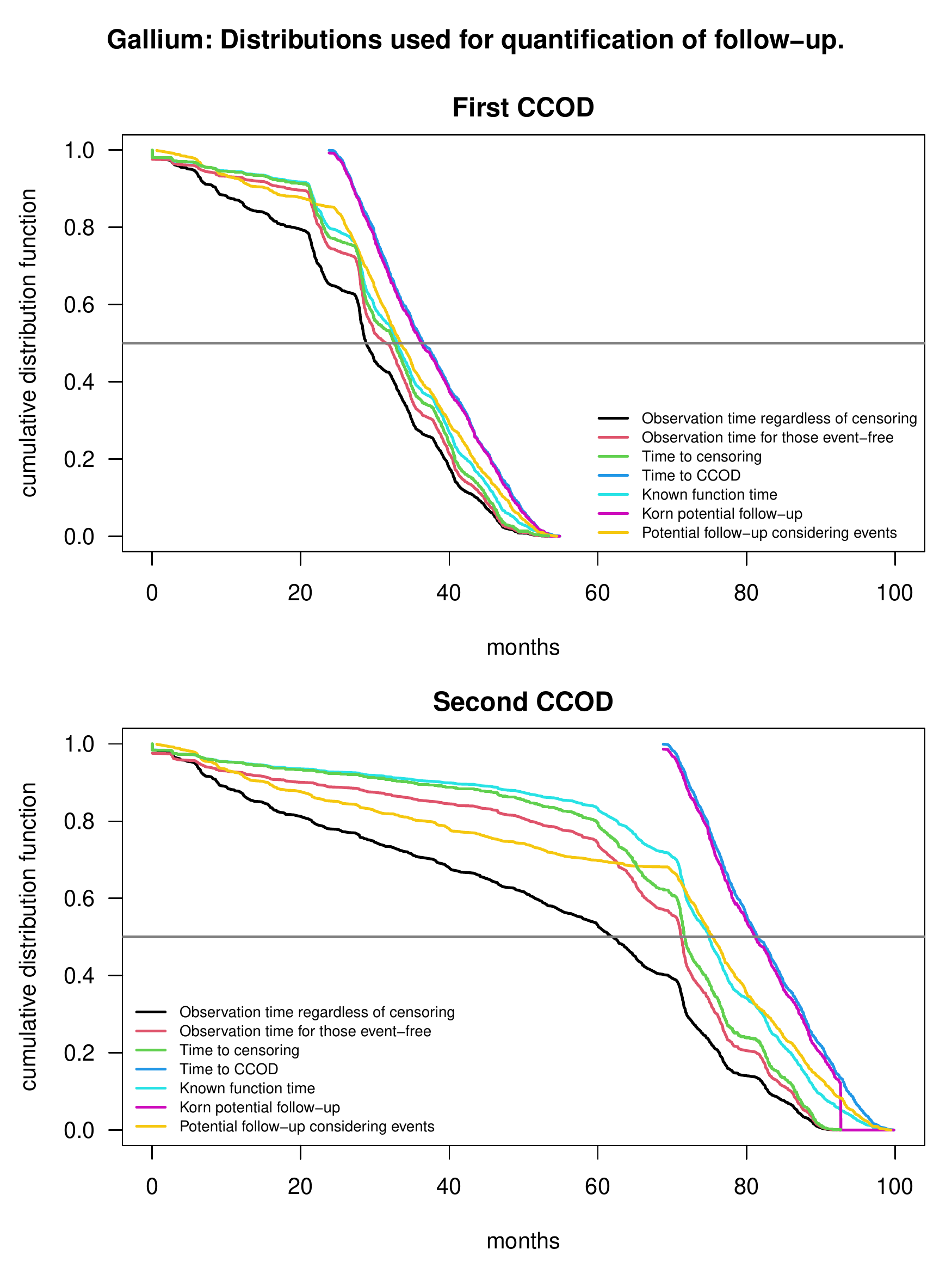}
\caption{Distributions from which follow-up quantifiers are derived, for Gallium.}
\label{fig:gallium dists}
\end{center}
\end{figure}

Figure~\ref{fig:gallium dists} provides the distributions corresponding to the quantities in Table~\ref{tab:quantities} from which the medians in Table~\ref{tab:gallium} were computed. The figure again illustrates the heterogeneity with which these various measures quantify follow-up. We also see that those quantities that involve the CCOD (\#4-\#7) in their computation exhibit a quite linear pattern, at different values for PFS. This simply illustrates that they all essentially map the accrual at least for a portion of the patients.

% -----------------------------------------------------
\section{Example with non-proportional hazards: delayed separation}
\label{ex:delayed}
% -----------------------------------------------------

Our second example is a trial with delayed separation. We simulated data for this trial (using the \code{R} package \code{rpact}, \cite{rpact}) making the following assumptions for a T2E endpoint such as PFS or OS:
\bi
\item The base event rate is 0.012, corresponding to a median T2E of 60 months.
\item A piecewise exponential survival with no effect between 0 and 12 months, and a HR of 0.65 thereafter.
\item In both arms, the probability that a patient is LTFU follows an exponential distribution calibrated such that this probability amounts to 0.025 at 12 months. This corresponds to a median time-to-LTFU of 329 months.
\item After a ramp-up of 6 months we recruit 42 patients per month until a maximal number of 1000 patients.
\ei

\begin{figure}[h!tb]
\begin{center}
\setkeys{Gin}{width=1\textwidth}
\includegraphics{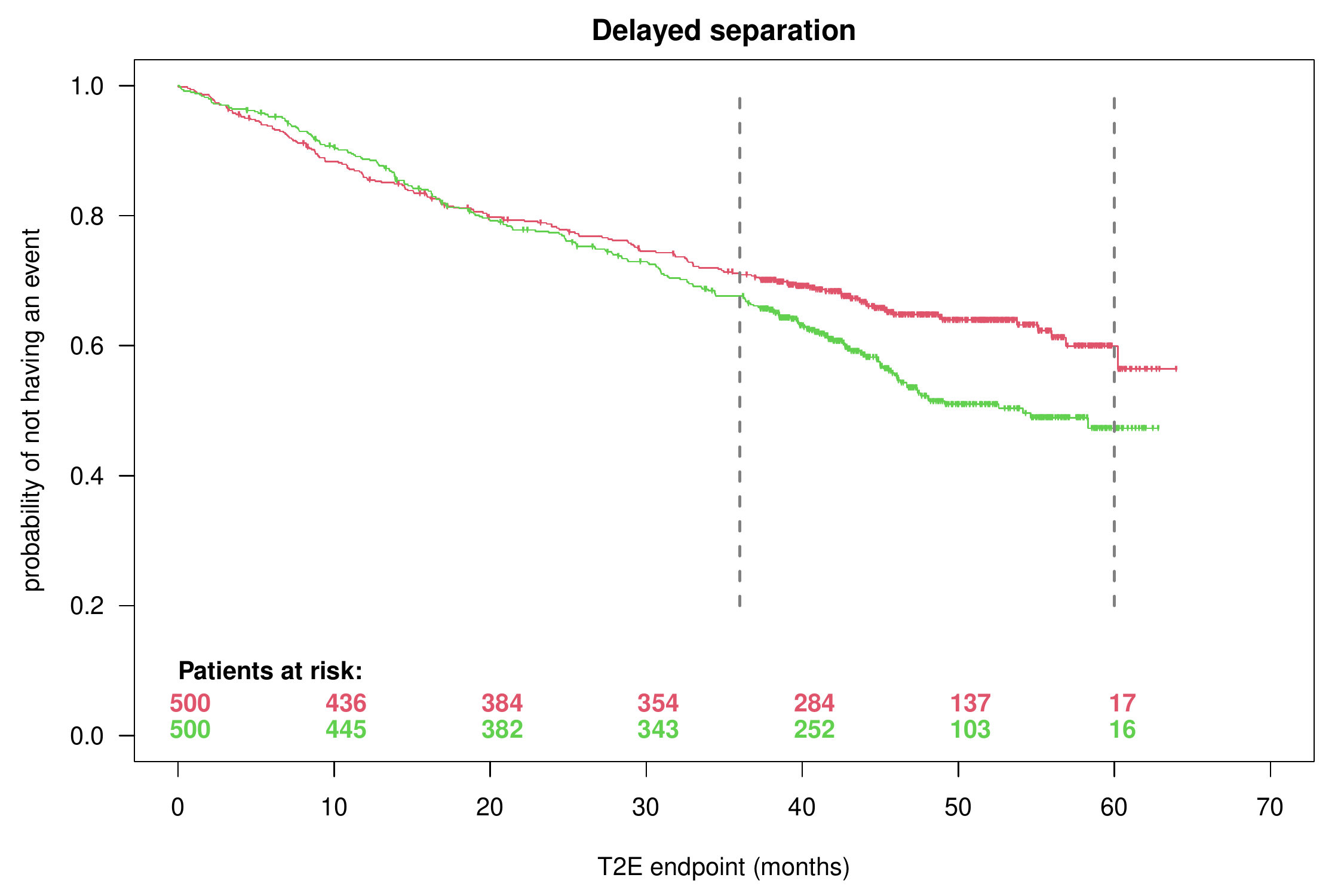}
\caption{T2E endpoint for Example 2. Vertical lines at milestone timepoints 36 and 60 months.}
\label{fig:ex2}
\end{center}
\end{figure}

With these assumptions a CCOD after 389 events gives a power of 80.5\% to reject the null hypothesis of no effect using an unweighted logrank test and 69.7\% power using the RMST difference based on \km estimates between arms, with a data-driven restriction time $t_0$ of the lower of the two maximal observed times (events and censored) in each arm. Both power values are based on 10000 simulated trials. Figure~\ref{fig:ex2} gives \km estimates of one such simulated trial.

70 patients (7.0\%) are LTFU and 541 patients (54.1\%) administratively censored.

Note that we only look at one CCOD for this example, as no additional insights appeared to be gained from having two beyond what we have already discussed for the first example.

% -----------------------------------------------------
\subsection{Answering the scientific questions}
\label{delayed:q}
% -----------------------------------------------------

% -----------------------------------------------------
\paragraph{Question 5: Precision.}
\label{delayed:q5}
% -----------------------------------------------------

Compared to effect quantification in the case of proportional hazards in Section~\ref{gallium:q5} an effect measure different than the HR may be indicated in this scenario. One option is the difference of RMST between arms, with an estimate of 2.82 months between arms with 95\% CI from -0.35 to 5.98. Estimated survival probabilities at clinically meaningful milestones provided in Table~\ref{tab:delayed_milestones} are also useful to describe the effect and precision for delayed separation.

\begin{center}
\renewcommand{\baselinestretch}{1.1}
% latex table generated in R 4.2.2 by xtable 1.8-4 package
% Mon Mar 13 23:13:02 2023
\begin{table}[h]
\centering
\begingroup\normalsize
\begin{tabular}{ccc}
  \hline
milestone & treatment arm & KM estimates and 95\% CIs \\
  \hline
 & Control arm & 0.68 [0.62, 0.73] \\
  36
months & Treatment arm & 0.71 [0.66, 0.76] \\
   & Difference treatment - control & 0.04 [-0.03, 0.09] \\
   \hline
 & Control arm & 0.47 [0.30, 0.65] \\
  60
months & Treatment arm & 0.60 [0.46, 0.73] \\
   & Difference treatment - control & 0.13 [0.05, 0.21] \\
   \hline
\end{tabular}
\endgroup
\caption{Milestone estimates for delayed separation example.}
\label{tab:delayed_milestones}
\end{table}\renewcommand{\baselinestretch}{1}
\end{center}

% -----------------------------------------------------
\paragraph{Question 6: Stability.}
\label{delayed:q6}
% -----------------------------------------------------

\begin{figure}[h!tb]
\begin{center}
\setkeys{Gin}{width=1\textwidth}
\includegraphics{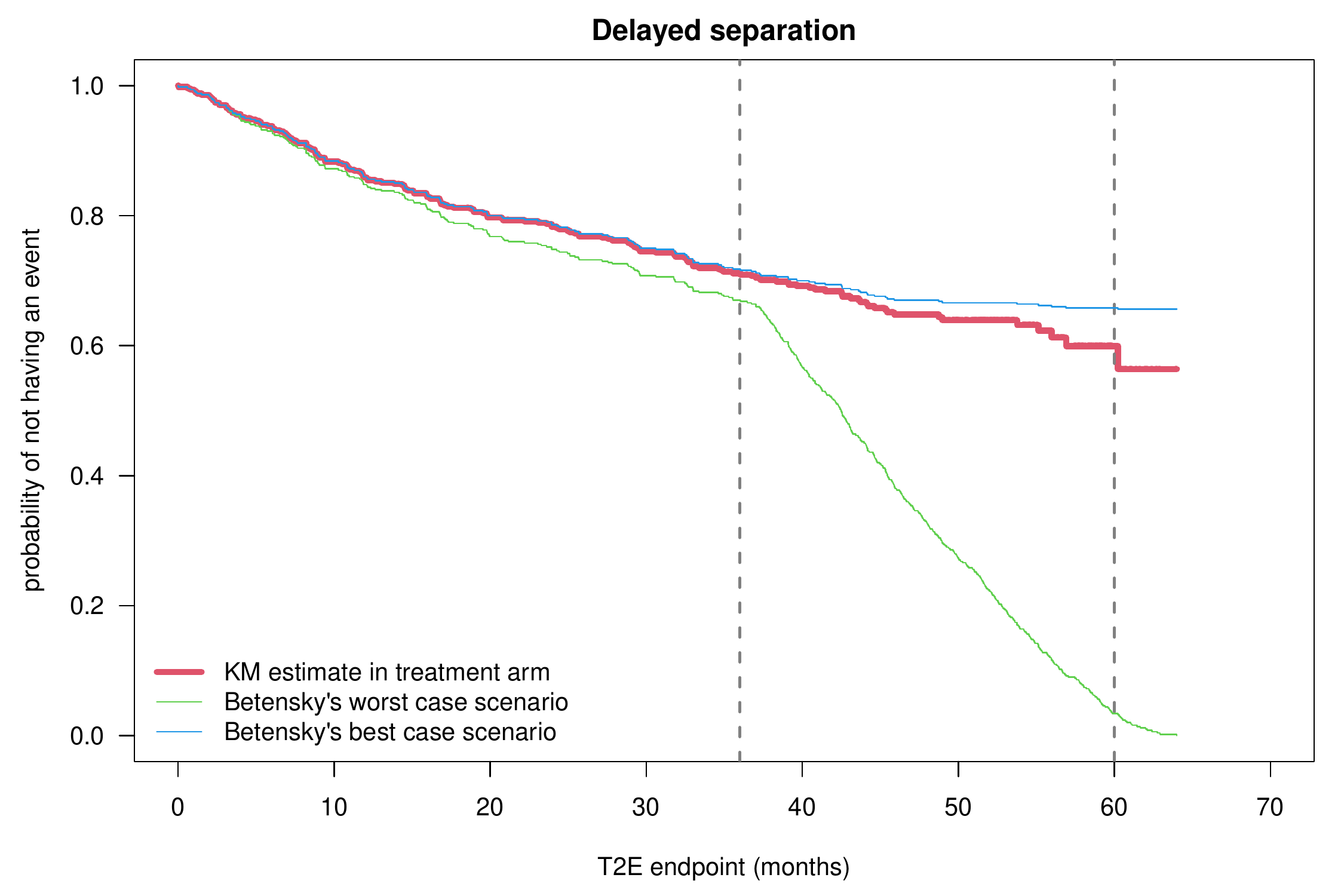}
\caption{Betensky's stability scenarios for the treatment arm of the "delayed separation" example.}
\label{fig:delayed_betensky}
\end{center}
\end{figure}

Both, RMST and milestone probabilities, typically rely on the \km estimate and for them to be valid effect quantifiers no further assumptions (apart from random censoring) are either required or typically made. Without such further assumptions it is only possible to construct rather crude bounds using the properties of the \km estimator and the stability principle proposed by Betensky for a \km estimate of a survival function: events will remain events also in future snapshots, but for currently censored observations two things can happen:
\bi
\item Best case scenario: every censored patient is censored at the last observed event date.
\item Worst case scenario: every censored patient has event day after at censoring date.
\ei
We exemplify this for the treatment arm in Figure~\ref{fig:delayed_betensky}.

Figure~\ref{fig:delayed_betensky} can be interpreted as follows: if we had complete observations, i.e. no censored patients, no amount of follow-up would change our \km estimate and the lower and upper extreme scenario bound would coincide. With censored observations the green and blue functions in the figure are the limits of where the \km estimate can go with more follow-up. Betensky proposes to use the RMST difference between the two extreme functions, normalized by the largest observed event-time, as a one-number measure of stability that ranges from 0 (complete stability) to 1 (complete instability). We compute 18.0\% in our example in the treatment and 16.1\% in the control arm.

In theory, we could now take the upper limit for the treatment and lower limit for the control arm and compute the RMST difference between these two functions. That would give the deterministically most extreme possible RMST value that is still possible. However, we do not consider this number adding much value to the discussion of stability.

To wrap this paragraph up, again, no quantification of follow-up whatsoever can provide information on stability, considerations as above appear to be more informative.

% -----------------------------------------------------
\paragraph{Question 7: Information.}
\label{delayed:q7}
% -----------------------------------------------------

The most important aspect is that for non-PH, number of events are typically not sufficient to quantify the "information" present at a given snapshot. Rather, information -- understood as the inverse of the variance of the group difference parameter -- typically depends on many more estimated quantities. We refer to the respective paragraph in Section~\ref{q2_noph} for a discussion of "information" in absence of PH, applied to RMST.

% -----------------------------------------------------
\paragraph{Question 8: Reliability.}
\label{delayed:q8}
% -----------------------------------------------------

An assessment of PH is not indicated in case of non-PH.

% -----------------------------------------------------
\paragraph{Question 9: Comparison of censoring distributions.}
\label{delayed:q9}
% -----------------------------------------------------

In this simulated dataset with assumed identical recruitment and censoring pattern in both arms censoring distributions are identical. We therefore omit the corresponding figure.

% -----------------------------------------------------
\subsection{Quantification of follow-up}
\label{delayed:qfu}
% -----------------------------------------------------

\begin{figure}[h!tb]
\begin{center}
\setkeys{Gin}{width=1\textwidth}
\includegraphics{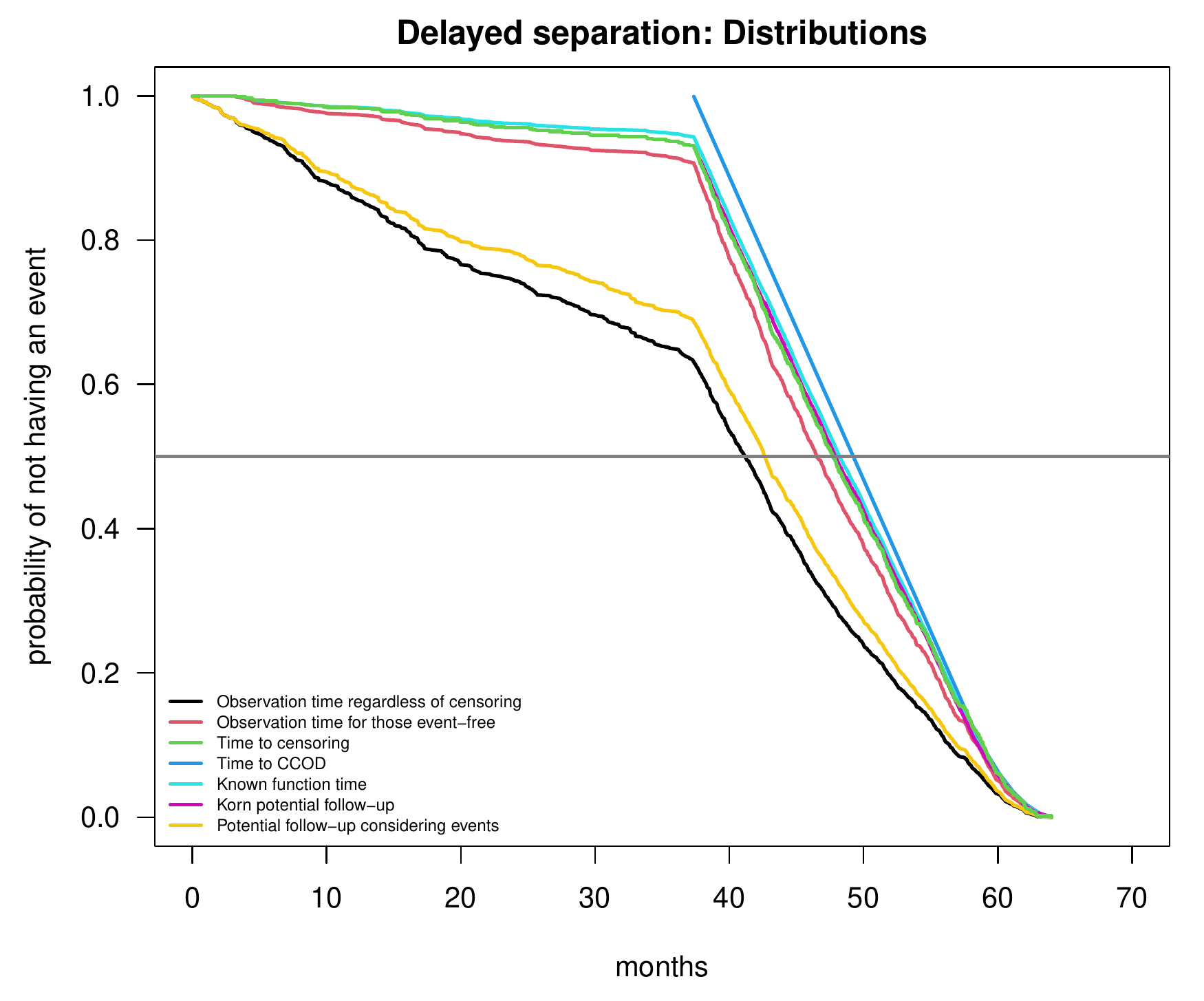}
\caption{Distributions from which follow-up quantifiers are derived, for the delayed separation example.}
\label{fig:nph dists}
\end{center}
\end{figure}

Figure~\ref{fig:nph dists} again provides the distributions from which the quantities in Table~\ref{tab:quantities nph} are derived. With values from 41.2 to 49.3 the heterogeneity in numbers is comparable to our first example.

\begin{center}
\renewcommand{\baselinestretch}{1.1}
% latex table generated in R 4.2.2 by xtable 1.8-4 package
% Mon Mar 13 23:13:03 2023
\begin{table}[h]
\centering
\begingroup\normalsize
\begin{tabular}{lc}
  \hline
 & CCOD 1 \\
  \hline
Observation time regardless of censoring & 41.2 \\
  Observation time for those event-free & 46.5 \\
  Time to censoring & 47.7 \\
  Time to CCOD & 49.3 \\
  Known function time & 48.2 \\
  Korn potential follow-up & 47.9 \\
  Potential follow-up considering events & 42.7 \\
   \hline
\end{tabular}
\endgroup
\caption{Different quantifications of follow-up for delayed separation example, in months.}
\label{tab:quantities nph}
\end{table}\renewcommand{\baselinestretch}{1}
\end{center}

% -----------------------------------------------------
\section{Further aspects}
\label{further}
% -----------------------------------------------------

Often, trialists insist on "enough follow-up for safety". While this is of course an important consideration it is an openly formulated question and again, no summary statistic of whatever follow-up quantity is helpful in deciding whether we have "enough follow-up for safety". Instead, if collection of safety data should inform length of follow-up, scientific questions we have for safety aspects of the interventions in the trial need to be formulated and estimands aligned to it defined. These then inform the data collection strategy.

When discussing and comparing results from multiple trials follow-up is often directly compared. Given that virtually never a definition of follow-up is given, this bears the high risk of comparing apples to oranges. Below, we anyway recommend to dispense with any single measure of follow-up (which would make a concern about inaccurate cross-trial comparisons obsolete) and if provided nevertheless, at least clearly define the considered quantities. Ideally, trialists should focus on the questions and quantities of interest to describe efficacy, as summarized in Tables~\ref{tab:examples1}-\ref{tab:examples2}, and compare these between trials, not some unclearly defined measure of follow-up.

Of note, such cross-trial comparisons are often made in Health Technology Assessments of drugs, when assessing whether the evidence base is suitable for evidence synthesis. Comparison of median follow-up time quantities is often part of such feasibility assessments. Again, relying on the true quantifiers of efficacy instead of unclearly defined follow-up, or at least clearly defining the latter, would make such assessments more robust and scientifically valid.

A reviewer commented on the potential for Bayesian methods in this context. Of course, wherever we estimate a clearly defined quantity we can use Bayesian equally well as frequentist approaches.

% -----------------------------------------------------
\section{(Ir)relevance of quantification of follow-up and recommendations}
\label{irrelevance}
% -----------------------------------------------------

In Tables~\ref{tab:examples1}, \ref{tab:examples2}, and \ref{tab:examples2} we made an attempt to list all questions trialists typically have for a trial with a T2E primary endpoint, be it for one or two arms and with potential follow-up snapshots. We also provided illustrative examples. Remarkably, all these questions could be answered without reference to any kind of follow-up whatsoever. So, we concur with Shuster\cite{shuster_91}  who already wrote

\begin{quote}
{\it ...it is recommended that fixed-term \km estimates and SEs be used to describe a quick numeric summary of outcome, rather than median followup. It is believed by this writer that many clinical investigators were unaware of the fact that the \km function adjusts for variable lengths of follow-up and provides an unbiased estimate of the true target population survival function, provided that competing losses are uninformative. Hence, they invented a term to meet a perceived need that, in fact, does not exist.}
\end{quote}

In our opinion, this conclusion still stands thirty years later. We have tried to make the questions that trialists ask for a trial with a T2E endpoint precise and discuss how these can be answered. No imprecisely and heterogeneously defined single "median follow-up" number is needed or useful to answer these questions and we therefore recommend not to report any at all. Instead we propose the following when reporting results of a T2E endpoint in a SAT or RCT:
\bi
\item Be clear on the scientific questions you want to answer.
\item It should be made clear to anyone involved in trials with T2E endpoint that there is no hope that a single number, however defined, can say everything about "follow-up", answer all the relevant questions trialists have, or allow to compare relevant aspects of trials.
\item The primary goal of the trial is an assessment of efficacy, so estimate(s) of $S_X$ in all treatment groups should be given.
\item As part of the assessment of efficacy, discuss precision, stability, information, and potential assumptions {\it separately} for any quantity of interest, i.e. typically estimates of survival functions and, in case of a two-arm trial, effect measures. We have exemplified this in our examples.
\item As discussed in our examples, estimates of $S_C$ for each treatment arm are very informative as well.
\item To describe accrual stating of $T_1$ and $T_2$ may be sufficient if accrual was approximately uniform as in Figure~\ref{fig:gallium_accrual}, as Altman et al.\cite{altman_95_surv} already recommend. If the recruitment pattern was not uniform the CDF of the accrual may be very informative.
\item The CCOD $T_3$ should be given. %Often, it is also informative to give the date when the last patient finished treatment.
\item Depending on the distribution of censoring reasons (administrative vs. LTFU) it might also be informative to describe these. A distinction may even enter Betensky's stability assessment in Figure~\ref{fig:delayed_betensky}, as patients LTFU are known to remain censored forever.
\item An option to complement, but not replace, the assessment of precision is to add patient numbers still at risk below the \km plot as we have done in Figures~\ref{fig:gallium} and \ref{fig:ex2}, and as recommended by Morris et al.\cite{morris_19}. Again, a split into the different censoring reasons might add additional information. One could even consider using different colors by censoring reason for the censoring ticks in the \km estimate. Furthermore, having these numbers may inform from which timepoint on the reader may want to start ignoring the \km estimates, even in absence of confidence intervals or bands.
\item We do not consider it necessary to provide any number that quantifies follow-up - it is simply not clear what questions it answers that cannot be more directly answered with other means. If trialists still want to give such a number then at least very clearly define how it is computed, in line with the recommendations by Altman et al.\cite{altman_95_surv}, later reiterated by Betensky\cite{betensky_15}. In line with Schemper et al.\cite{schemper_96} we consider time-to-censoring, i.e. an estimate of $S_C$, to be most informative. This also allows an assessment whether the censoring pattern is different between arms, as in Figure~\ref{fig:gallium_cens_arm}. Reporting only the estimated median has the usual deficiency of summarizing an entire distribution into just one number, which likely will only give an incomplete account of what is actually happening.
\item Key differences in case of non-PH are: (1) an effect quantifier other than the HR and associated hypothesis test must be chosen, (2) because of the absence of PH the assessment of stability of the effect estimate becomes more important and (3) information, and as a consequence the power of any hypothesis test, does depend on (many) more quantities than just the number of events. Which ones exactly depends on the effect measure that is used. Valid estimates of the censoring distributions are therefore even more relevant in this case.
\ei

With these recommendations we hope to provide guidance to trialists what scientific questions are relevant in connection with trial follow-up, how to best analyze follow-up and answer these questions, and to motivate them to {\it finally}, thirty years after Shuster's letter, move away from single-number quantifiers of follow-up.

% -----------------------------------------------------
\section{Code Availability Statement}
\label{software}
% -----------------------------------------------------

Code that implements all the methods discussed in this paper is available at \url{https://oncoestimand.github.io/quantFU/quantFU.html}.

% -----------------------------------------------------
\section{Data Availability Statement}
\label{sharing}
% -----------------------------------------------------

Qualified researchers may request access to individual patient level data of the Gallium trial\cite{marcus_17} through the clinical study data request platform \url{https://vivli.org}. Further details on Roche's criteria for eligible studies are available here: \url{https://vivli.org/members/ourmembers}. For further details on Roche's Global Policy on the Sharing of Clinical Information and how to request access to related clinical study documents, see here: \url{https://www.roche.com/research_and_development/who_we_are_how_we_work/clinical_trials/our_commitment_to_data_sharing.htm}.

% -----------------------------------------------------
\section{Acknowledgments}
\label{ack}
% -----------------------------------------------------

This paper has been written within the industry working group {\it estimands in oncology}, which is both, a {\it European special interest group ``Estimands in oncology'', sponsored by PSI and European Federation of Statisticians in the Pharmaceutical Industry (EFSPI)} and a scientific working group of the biopharmaceutical section of the American Statistical Association. Details are available on \url{www.oncoestimand.org}.

We thank Keaven Anderson, Sandro Gsteiger, Tina Nielsen, and Godwin Yung, as well as the associate editor and two reviewers for helpful comments.

\appendix

% -----------------------------------------------------
\section*{Appendix}
\label{sec:appendix}
% -----------------------------------------------------

\bibliographystyle{ama}
\bibliography{stat}

\end{document}